\documentclass[twocolumn,showpacs,floatfix,superscriptaddress,nofootinbib,amssymb,amsmath,aps,prc,eqsecnum]{revtex4}
\usepackage{graphicx}
\usepackage{bm}
\usepackage{amsmath}

\usepackage{amssymb}

\newcommand{\vn}[1]{V_{#1}}

\newcommand{\pt}{p_{t}}
\newcommand{\raa}{R_{AA}}
\newcommand{\rpa}{R_{pA}}
\newcommand{\psih}[1]{\psi_{#1}^{J}}
\newcommand{\psie}[1]{\psi_{#1}^{\mathrm{EP}}}
\newcommand{\psip}[1]{\psi_{#1}^{\mathrm{PP}}}

\begin{document}


\title{Jet Quenching and Its Azimuthal Anisotropy \\ in AA and possibly High Multiplicity pA and dA Collisions}

\author{Xilin Zhang} \email{zhangx4@ohio.edu}
\affiliation{Institute of Nuclear and Particle Physics and Department of
Physics and Astronomy, \\
Ohio University, Athens, OH 45701, USA.}

\author{Jinfeng Liao} \email{liaoji@indiana.edu}
\affiliation{Physics Department and Center for Exploration of Energy and Matter, \\
Indiana University, 2401 N Milo B. Sampson Lane, Bloomington, IN 47408, USA.} 
\affiliation{RIKEN BNL Research Center, Bldg. 510A, \\ Brookhaven National Laboratory, Upton, NY 11973, USA.}

\begin{abstract}
We report our calculation of jet quenching and its azimuthal anisotropy in the high energy AA and high multiplicity pA and dA collisions. The purpose of this study is twofold. First, we improve our previous event-by-event studies, by properly implementing $\pt$ dependence in the modeling. We show that, within the jet ``monography'' scenario featuring a strong near-Tc-enhancement  of jet energy loss, the computed high-$p_t$ nuclear modification factor $\raa$ and the harmonic coefficients in its azimuthal anisotropy $\vn{2,\,3,\,4}$, agree well with the available data from both Relativistic Heavy Ion Collider (RHIC) and Large Hadron Collider (LHC). Second, in light of current discussions on possible final state collective behavior in the high-multiplicity pPb and dAu collisions, we examine the implication of final state jet attenuation in such collisions, by applying the same model used in AA collisions and quantifying the corresponding $\rpa$ and $R_{dA}$ and their azimuthal anisotropy. The high-$p_t$  $\vn{n}$ are an set of clean indicators of final state energy loss. In particular we find in the most central pPb (5.02 TeV) and dAu (200 GeV) collisions, $\vn{2}$ is on the order of $0.01$ and $0.1$ respectively, measurable with current experimental accuracy. In addition, our high-$\pt$ $R_{dA}$ is around $0.6$ which is compatible with preliminary dAu results from RHIC.       
\end{abstract}

\maketitle

\section{Introduction} \label{sec:intro}

In the high energy heavy ion collisions, a new state of strongly interacting matter, the quark-gluon plasma (QGP) is created. Measuring the properties of this matter is of fundamental interest. Such measurements are now done at a variety of collisional beam energies at both RHIC and LHC. 

The highly energetic partons (which ultimately turn into observed jets) born from the very initial hard scatterings provide an invaluable way of probing the matter properties. Along the way penetrating the medium, a jet parton will interact with the medium and lose energy, which can be  manifested through comparing the observables (such as leading hadron yield) in the jet kinematic region with the same observables measured from the reference pp collisions.  One commonly used observable in jet quenching study  is the nuclear modification factor, $\raa$, defined as the ratio between the hadron production in AA collision and that in NN collision (further scaled by the expected binary collision number). This observable turns out to be a crucial quantity in the establishment of QGP: a significant suppression of high-$p_t$ (transverse momentum) hadron production was first observed at RHIC and then at LHC \cite{LHC_suppression},  with $R_{AA}$ reaching  $\sim 0.2$ in the most central collisions. Measurements of charged particle, identified hadron, heavy flavor, and photon production have coherently pointed to a medium that is extremely opaque to {\em ``colored''} hard probe.  For reviews see e.g. \cite{Gyulassy:2003mc}. 

Another important aspect of jet-quenching analysis, the so-called geometric tomography \cite{Gyulassy:2000gk,Wang:2000fq}, is about the azimuthal anisotropy of jet energy loss. The geometry of the hot medium created in AA collision event is generally anisotropic in the transverse plane (almond-like), and so is the high-energy parton energy loss moving along different azimuthal angle ($\phi$), which leads to a measurable anisotropy in the $R_{AA}(\phi)$ \cite{Shuryak:2001me,Liao:2008dk,Jia:2010ee,Jia:2011pi,Liao:2011kr}. More recently this idea has been extended to systematically quantify the jet response to the the initial state fluctuations giving rise to various other harmonic coefficients in the high-$p_t$ region  \cite{Bass:2008rv,Rodriguez:2010di,Zhang:2012mi,Zhang:2012ie, Zhang:2012ha,Jia:2012ez,Betz:2011tu}.

Detailed measurements of jet quenching and its azimuthal anisotropy via leading hadrons have been done over the past decade, with large amounts of data about how $\raa$ depends on hadron's $p_t$ and azimuthal angle $\phi$, the collision centrality class, and  the beam energy. It is essential to describe all these characteristics within a unified jet quenching modeling scheme. The first goal of the present paper is to improve our previous studies \cite{Liao:2011kr,Zhang:2012mi,Zhang:2012ie,Zhang:2012ha} on this subject, and to perform comprehensive computations of the mentioned dependences and compare the results with available data. The jet energy loss model we use here is based on the jet ``monography'' scenario featuring a strong  near-Tc-enhancement (NTcE) of jet-medium interaction.  This scenario ~\cite{Liao:2008dk} was first discovered in the effort to explain the large jet quenching anisotropy (\'{a} la $\vn{2}$ at high $p_t$) at RHIC~\cite{star2005,phenix2010,Jia:2010ee,Jia:2011pi}, and is based on the ``magnetic scenario'' for sQGP \citep{Liao:2006ry, Liao:2008vj, Chernodub, Ratti:2008jz,D'Alessandro:2010xg,Bornyakov:2011dj}. More recently it was shown in Refs.~\cite{Zhang:2012ie,Zhang:2012ha} that the NTcE model naturally induces a reduction ($\sim 30\%$) of jet-medium interaction strength from RHIC to LHC. Such reduction is required to explain both RHIC and LHC $\raa$ consistently according to the analysis in Ref.~\cite{Betz:2012qq}. In our previous works, the NTcE model has been shown to well describe the centrality trend, the anisotropy, and the beam energy dependence of jet quenching. However  we've used the fractional energy loss  (largely motivated by the ``flatness'' seen at RHIC) leading to no $p_t$-dependence for the observables. In this study, we  introduce the $\pt$ dependence into our jet quenching modeling in a similar way as used in Refs.~\cite{Betz:2011tu,Horowitz:2011gd}. As a result, we can compute the jet quenching observables in different $p_t$ bins, which is needed to analyze the jet quenching data from LHC.  
 
The second goal of this study is to examine possible final state jet-medium interaction in the high multiplicity pA and dA collision events. Such events have generated a great deal of interests recently.  A  di-hadron correlation pattern that is long-range in rapidity while collimated in azimuthal angle (also called as ``ridge'') has been observed in high-multiplicity proton-proton (pp at $\sqrt{s_{NN}}=7$ TeV) \cite{Khachatryan:2010gv},  proton-lead (pPb at $\sqrt{s_{NN}}=5.02$ TeV) \cite{CMS:2012qk,Abelev:2012cya,Aad:2012gla,Aad:2013fja,ALICE:2012mj}, and deuteron-gold collisions (dAu at $\sqrt{s_{NN}}=200$ GeV) \cite{Adare:2013piz}.  The same correlation has been regarded as an important indicator of the collective behavior of the matter created in AA collisions at RHIC \cite{Abelev:2009af} and LHC \cite{Chatrchyan:2011eka,ATLAS:2012at}. Previous AA studies  have concluded that due to the initial state geometry and its fluctuations, different harmonic flows are developed during matter's expansion, which upon folding together create the observed ``ridge'' correlation \cite{Alver:2010gr,Alver:2010dn,Luzum:2011jpg,Sorensen:2008bf,Teaney:2010vd,Qiu:2011iv,Staig:2010pn,Takahashi:2009na,Schenke:2010rr,Xu:2010du,Qin:2010pf,Ma:2010dv}. The question still under debate, is whether the ``ridge'' in pp, pPb, and dAu collisions implies the formation of a thermal QGP in these small systems. Efforts to understand the origin of the ``ridge'' in these collisions are still ongoing \cite{Li:2012hc}. The hydrodynamic calculations (assuming enough final state interaction to make a nearly thermal medium) have been carried out for  pp \cite{Deng:2011at,Shuryak:2010wp}, pPb, and dPb collisions \cite{bozek2012prc,bozek2013,Bozek:2013df}. At the same time, there are alternative explanations based on possible intrinsic azimuthal correlations in the early gluon production~\cite{Levin:2011fb,Dusling:2012iga}. 

While it seems plausible that collective effect is negligible in the pp case,  certain degrees of collectivity could be present in the dense partonic system  created in the pA and dA collisions (which may be tentatively called ``mini-bang''). 
To approve or disapprove this assertion, it is crucial to have independent probes in addition to the soft-sector observables (e.g. flows, soft ``ridge''). We suggest to use the final state  interaction induced jet energy loss as an indicator. In this paper, we will make a first analysis of jet quenching and its azimuthal anisotropy in these ``mini-bangs'' by applying our energy loss modeling (calibrated in AA collisions) and computing $R_{pA}$ and $R_{dA}$, and their azimuthal anisotropy's harmonic coefficients $\vn{1,2,3}$.  It is worth mentioning here that the computed $R_{dA}$ compares favorably with recent dAu results \cite{Adler:2006wg,Sahlmueller:2012ru,Perepelitsa:2013jua}. 
In particular as we will show, the high-$\pt$ anisotropy, not easily contaminated by initial state effects, can serve as a ``clean'' signal of final state jet-medium interaction. (Note there are recent theoretical works \cite{Zakharov:2013lga,Zakharov:2013gya} about jet quenching in the high multiplicity pp collisions.)  

The rest of the paper is organized as follows. Sec.~\ref{sec:NTcEmodel} is devoted to the details of the NTcE model improved by the implementation of $\pt$ dependence. We use central $\raa$ to fix two model parameters, and then compute $\raa$ in different centrality classes and $\pt$ bins for both RHIC and LHC collisions.  In Sec.~\ref{sec:anisotropyAA}, we extract the harmonic coefficients $\vn{2,3,4}$ from $\raa$'s azimuthal anisotropy. In Secs.~\ref{sec:ppb} and~\ref{sec:dAu}, we explore possible jet quenching and its anisotropy in the pPb and dAu collisions and make predictions for future measurements. Finally we summarize the results in Sec.~\ref{sec:sum}.  

\section{Jet quenching in AA collisions from RHIC to LHC} \label{sec:NTcEmodel}

In this work, we use the geometric model scheme  \cite{Shuryak:2001me,Liao:2008dk,Jia:2010ee,Jia:2011pi,Betz:2011tu,Betz:2012qq,horwitz2011,Liao:2011kr,Zhang:2012mi,Zhang:2012ie,Zhang:2012ha}, in which the geometric features of jet energy loss (e.g. the path-length dependence) are implemented in a phenomenological way. In this class of models, the differential energy loss ($dE/dl$) experienced by a jet parton through the medium is expressed as 
\begin{eqnarray}
dE =  - E^\delta \,  \kappa[s(l)]\, s(l)\, l^m \, d \, l \ . \label{eqn:jeteloss1} 
\end{eqnarray} 
Here $s(l)$ is the local entropy density along the jet path, and $\kappa(s)$ is the local jet quenching strength which as a property of matter should in principle depend on $s(l)$.
There are different choices of $m$ and $\kappa(s)$ \cite{Jia:2010ee,Jia:2011pi,Betz:2011tu,Betz:2012qq,horwitz2011}. We only focus on the mentioned NTcE model which assumes $m=1$ based on the LPM effect in the radiative energy loss \cite{Baier:1996sk}, and introduce a strong jet quenching component in the vicinity of $T_c$ (with density $s_c$ and span of $s_w$) via 
\begin{eqnarray} \label{Eq_kappa}
\kappa(s)=\kappa [1+ \xi\, \exp(-(s-s_c)^2/s_w^2)] \ . 
\end{eqnarray}
The parameters are $\xi=6$, $s_c=7/fm^3$, and $s_w=2/fm^3$. As demonstrated previously in Refs.~\cite{Liao:2008dk,Liao:2011kr,Zhang:2012ie,Zhang:2012ha}, this model reproduces the high $\pt$ $V_{2}$ at RHIC, and  naturally explains the nontrivial reduction of the jet-medium interaction strength (in average) from RHIC to LHC. %

In these previous works~\cite{Liao:2008dk,Liao:2011kr,Zhang:2012ie,Zhang:2012ha}, the fractional energy loss scenario [i.e., setting $\delta=1$ in Eq.~(\ref{eqn:jeteloss1})] was used, motivated by the weak $p_t$-dependence of the RHIC data in the jet region. As a result the computed jet quenching observables are all $p_t$-independent. Such a model fails to describe the LHC $\raa$ data which show a strong $p_t$-dependence in a significantly larger jet $p_t$ region. Here we improve the model by treating the $\delta$ as a free parameter that needs to be fixed in order to reproduce $\raa$'s $\pt$-dependence at both RHIC \cite{star2003, phenix2008} and LHC \cite{cms2012,alice2012}.  

Let's define the ratio between initial and final jet energy as ${\mathcal R}_{\bf P,Ev}$ in a given AA collision event ${\bf Ev}$ for a particular jet path ${\bf P}$, i.e., $E_i = E_f {\mathcal R}_{\bf P,Ev}(E_f)$. Using Eq.~(\ref{eqn:jeteloss1}), we can get ($\delta \neq 1$):
\begin{eqnarray} 
{\mathcal R}_{\bf P,Ev} ( E_f )  &=& \left [ 1 + \frac{{(1-\delta) \mathcal D}_{\bf P,Ev}}{E_f^{1-\delta}} \right]^{\frac{1}{1-\delta}} \ , \notag \\
{\mathcal D}_{\bf P,Ev} &\equiv&   \int_{\bf P,Ev} \kappa[s(l)] s(l) l^m dl \ . \label{eqn:jeteloss2}
\end{eqnarray}
We focus on mid-rapidity measurements so that $E\approx p_t$. Suppose the reference pp spectrum at the same collision energy is $g_{pp}(\pt)$, based on Eq.~(\ref{eqn:jeteloss2}) we can get the formula for the averaged $\raa$ over many events:
\begin{eqnarray}
 &&R_{AA}(p_t,\phi) = \notag \\
 &&\frac{  < \, <   g_{pp}[p_t\, {\mathcal R}_{\bf P,Ev}\, ( p_t)] \, [{\mathcal R}_{\bf P,Ev}(p_t)]^{1+\delta} \,    >_{\bf P}  \,  >_{\bf Ev} }{g_{pp}(p_t)} \ . \notag \\
\end{eqnarray}
Here, $<<>_{\bf P}>_{\bf Ev}$ means averaging all the possible jet paths from different jet spots in each event and then averaging all the events.
The $p_{t}$ spectrum of pp collision at 200 GeV is expressed as $g_{pp}(p_{t}) \propto 1/{p_{t}^{n}}$ in the high-$p_t$ region.
In Ref.~\cite{horwitz2011}, a $\pt$-dependent index $n$ is introduced. This is supported by the pQCD calculation and also provides an explanation for $R_{AA}$'s weak $\pt$-dependence at RHIC (200 GeV). For the $\sqrt{s}=2.76$ TeV and higher energy at LHC (e.g., $\sqrt{s}=5.02$ TeV pPb collision), we use the scaling law \cite{arleo2010,cms2011spectrum}: 
$\sqrt{s}^{n}g_{pp}(p_{t})=p_{0}\left[1+{x_{T}}/{p_{1}}\right]^{p_{2}}$, while $x_{T}\equiv 2p_{t}/{\sqrt{s}}$.
The relevant (energy independent) parameters $p_{1}$ and $p_{2}$ have been fitted to the measurement at $\sqrt{s}=2.07$ TeV (see Ref.~\cite{cms2012} for details).  

To study $\raa$'s centrality and $\pt$ dependences (not azimuthal anisotropy), we use the optical Glauber model \cite{miller2007} to generate initial state density profile. The relation between the initial entropy density and participant density (and collision density) can be found in Ref.~\cite{Zhang:2012ha}. [At LHC (2.76 TeV), the entropy density is roughly doubled from RHIC (200 GeV) case \cite{cshen2012}.] The total NN cross sections, $\sigma_{NN}^{inel}$, at these energies are $42$ and $62$ mb \cite{heinz2011,miller2007}. In addition, only the longitudinal Bjorken expansion (along the collision axis) \cite{bjorken} is included, while the transverse dynamics is frozen.  
\begin{figure*}
\includegraphics[height=170mm, width=80mm, angle=-90]{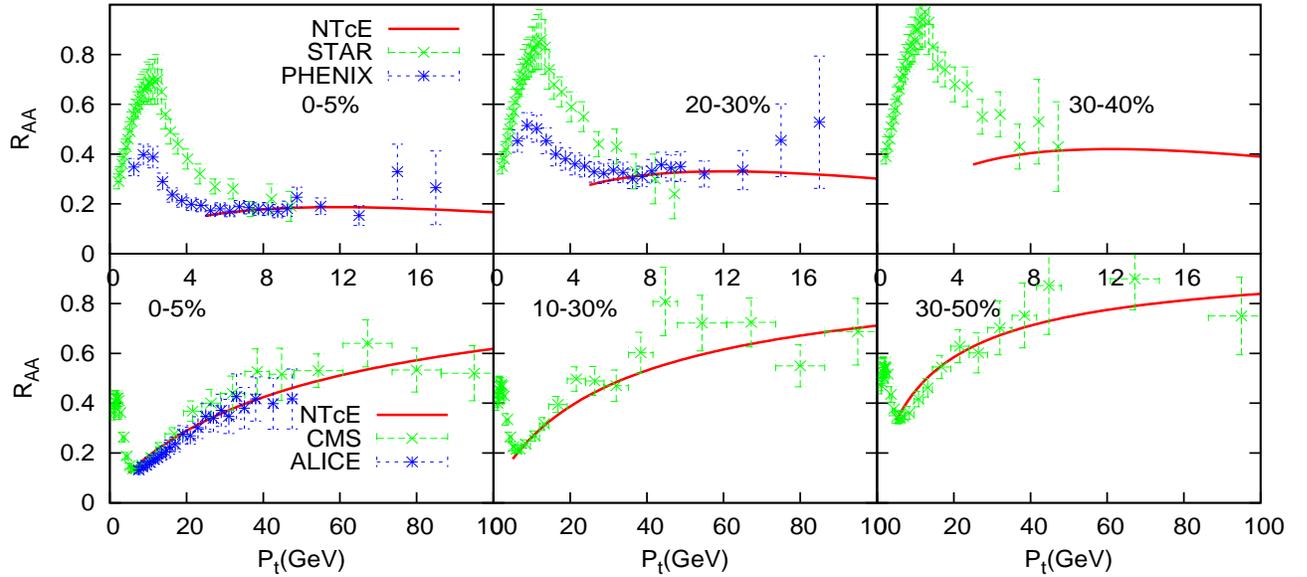}
\caption{$R_{AA}$ vs $\pt$ in different centrality classes of AA collisions at both RHIC and LHC. The data are from  Refs.~\cite{star2003, phenix2008,cms2012,alice2012}. } \label{fig:RAA}
\end{figure*}
To fix the parameters, $\delta$ and $\kappa$, in Eq.~(\ref{eqn:jeteloss1}), we use the most central ($0-5\%$) high-$\pt$ $\raa$ data from RHIC and LHC (see the plots for central collision in Fig.~\ref{fig:RAA}). A $\chi^2$-analysis has determined the optimal parameters to be: $\delta\approx 0.19$ and $\kappa\approx 0.0065$. (In Eq.~(\ref{eqn:jeteloss1}), energy unit is always GeV.)  
Then we calculate the $\raa$ at other centrality classes. The results for the two collision energies are collected in Fig.~\ref{fig:RAA}. The shown data are from Refs.~\cite{star2003, phenix2008,cms2012, alice2012}. As one can see, the present  model describes very well the $\raa$ in different centralities and $p_t$ bins.  The fact that the NTcE model gives correct medium opaqueness evolution from RHIC to LHC is nontrivial, as already emphasized in our previous analysis~\cite{Zhang:2012ie,Zhang:2012ha}. In this model, the RHIC fireball with its bulk matter staying closer and longer in the near-$T_c$ region is naturally more opaque than the LHC fireball, and we've quantified the reduction of the jet-medium interaction strength to be $\sim 30\%$ in this opaqueness evolution~\cite{Zhang:2012ha}. Such a reduction has been consistently reported from other studies~\cite{Betz:2012qq,Betz:2013caa,Buzzatti:2012dy}, where either a reduction of jet-medium coupling strength is put in by hand when extrapolating to LHC or a strong running coupling effect is introduced.

\section{High $\pt$ azimuthal anisotropy in AA collisions from RHIC to LHC} \label{sec:anisotropyAA}
To study $\raa$'s anisotropy, we use the Monte-Carlo (MC) Glauber model \cite{miller2007} to simulate the fluctuating initial conditions, and compute the azimuthal-angle $\phi$-dependent  $\raa(\phi,\pt)$ on an event-by-event basis. 
Except the initial state, all the other ingredients in this calculation are the same as those in the previous section. 
The details of implementing the MC-Glauber and a careful analysis of the initial state geometry (e.g., $\epsilon_{n}$) can be found in Ref.~\cite{Zhang:2012ha}. 
The medium is again assumed to experience only longitudinal Bjorken expansion \cite{bjorken}. One though may take caution that the transverse expansion in general will reduce the anisotropy of the matter density distribution and thus reduce the hard probe's responses to such anisotropy, as shown in e.g.~\cite{Betz:2013caa,Molnar:2013eqa}. A full modeling needs to combine the jet energy loss calculation with a realistic hydrodynamic modeling, which we are currently working on with the results to be reported in future.  
 
Due to the anisotropic and fluctuating matter geometry, the jet energy loss is different in each transverse direction, and for each event there will be an event-wise, azimuthal-angle $\phi$-dependent nuclear modification factor $\raa(\pt,\phi)$. Its anisotropy could be quantified by the harmonic coefficients $\vn{n}(\pt)$ from a Fourier decomposition: 
\begin{align}
 R_{AA}&(\phi,\pt) \equiv R_{AA}(\pt) \times \notag \\
&\left(1+2\sum_{n} \vn{n}(\pt)\cos\left[n(\phi-\psih{n}(\pt))\right]\right) \, . \label{eqn:Raadecomp}
\end{align}
Here we extract $V_{2}$, $V_{3}$ and $V_{4}$ in each event and the corresponding reference angles $\psih{2}$, $\psih{3}$, and $\psih{3}$ in different $\pt$ bins. The averaged $V_{2,3,4}$ can be directly compared with experimental data. 

\begin{figure*}
\includegraphics[height=170mm, width=80mm, angle=-90]{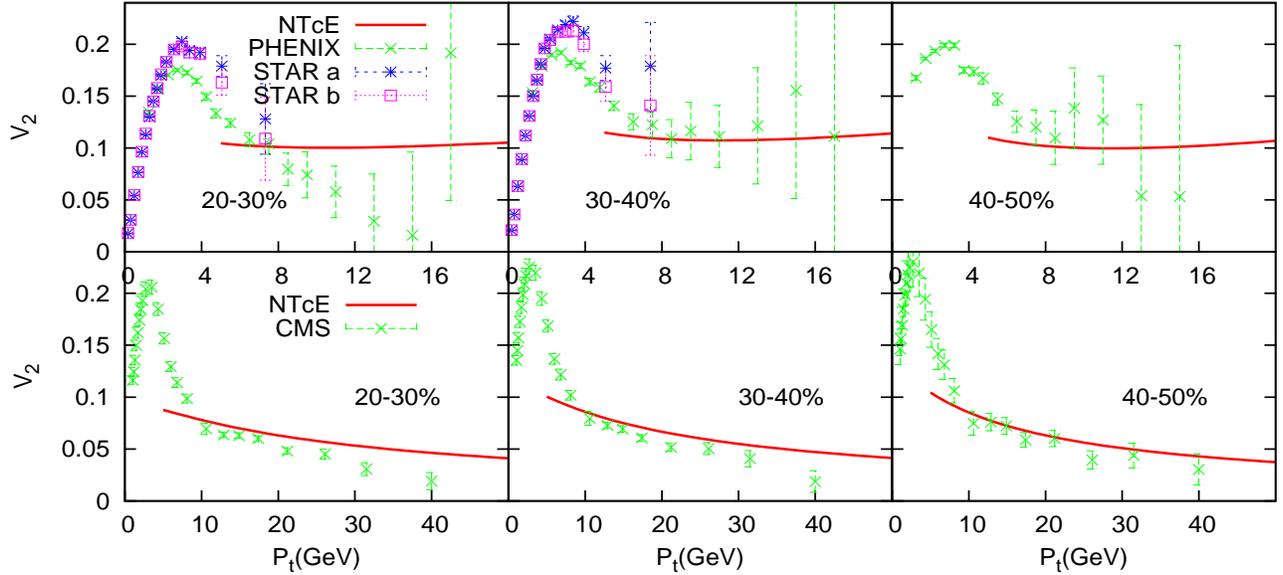}
\caption{$V_{2}$ vs $\pt$ in different centrality classes of AA collisions at RHIC and LHC. The data are from Refs.~\cite{star2005,phenix2010, cms2012azi}. In the upper panel, ``STAR a'' and ``STAR b'' are data based on different analysis methods~\cite{star2005}. } \label{fig:V2}
\end{figure*}
Let us first examine the second harmonics, $\vn{2}(\pt)$, at different centralities in Fig.~\ref{fig:V2}. The shown $\vn{2}(\pt)$ is the average value of all the event's $\vn{2}$ at a given $\pt$ bin. The data are from Refs.~\cite{star2005,phenix2010, cms2012azi}. Strictly speaking what   the experiments measure would be most close to the $\vn{2}$ relative to $\psie{2}$ (the event-plane of final soft hadrons), i.e. $\vn{2}\cos[2(\psih{2}-\psie{2})]$. Our study shows that $\psih{2}$ and $\psip{2}$ are well correlated: $<\cos[2(\psih{2}-\psip{2})]>$ is around $1$ in non-central collisions (see details in Ref.~\cite{Zhang:2012ha}) and has a very week $\pt$ dependence. In Ref.~\cite{ZhiQiu2011PRC84}, $\psie{2}$ and $\psip{2}$ are also shown to be strongly correlated. Because of the two strong correlations, the correction due to the mismatch among the different planes would be minor. It is thus sensible to compare our $\vn{2}$ with the experimental data for the non-central collisions. Also the impact of medium's transverse dynamics in the non-central collisions should be smaller than that in central ones because of the shorter jet path in average. We can see the NTcE model gives correct magnitude and $\pt$-dependence for $\vn{2}$ across different centralities at two collision energies. Similar to $\raa$,  the $\vn{2}$ at RHIC shows a weak $\pt$-dependence, while this dependence is rather strong at LHC. One does notice that for the LHC case, our model results are consistently higher than data: a plausible explanation could be the omitted transverse expansion effect which would reduce the anisotropy to certain extent and which is expected to be stronger at LHC.  

%
\begin{figure*}
\includegraphics[height=170mm, width=80mm, angle=-90]{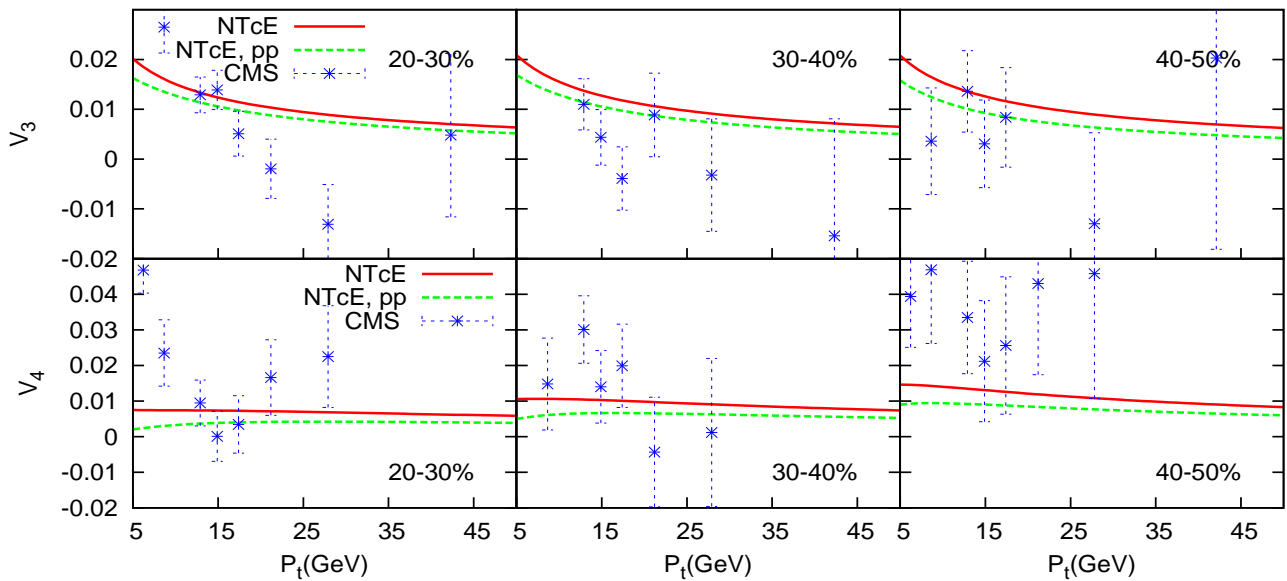}
\caption{$V_{3}$ vs $\pt$ (upper panel) and $V_{4}$ vs $\pt$ (lower panel) in different centrality classes of AA collisions at LHC. Two different results are shown here. ``NTcE'' gives the harmonic component $\vn{n}$, while ``NTcE pp'' is the $\vn{n}$ projected to the corresponding participant planes. See the discussion in text. The data are from  Ref.~\cite{CMSVn}.} \label{fig:V34}
\end{figure*}

Let us  move to the higher harmonics arising from fluctuations. The results of $\vn{3}(\pt)$ and $\vn{4}(\pt)$ at LHC are plotted in Fig.~\ref{fig:V34} in comparison with the available data from CMS \cite{CMSVn}. Two different results are shown in the plots: the ``NTcE'' results correspond to the averaged $\vn{n}$ ($n=3,\, 4$) from Eq.(\ref{eqn:Raadecomp}), while the ``NTcE, pp''  results are the averaged $\vn{n}\cos[n(\psih{n}-\psip{n})]$ ($n=3,\, 4$) which account for the mismatch between the initial participant plane angle and the final jet response plane angle. 
According to Ref.~\cite{ZhiQiu2011PRC84}, $\psie{3}$ and $\psip{3}$, similar to $\psie{2}$ and $\psip{2}$, correlate quite well in all the centrality classes, but the correlation between $\psie{4}$ and $\psip{4}$ is complicated and even changes the sign. Meanwhile by comparing ``NTcE'' and ``NTcE, pp'', we can see the correlation between $\psih{3}$ and $\psip{3}$ is also  stronger than that between $\psih{4}$ and $\psip{4}$.
Keeping in mind the difference between initial participant plane and the final event plane angles, it is nevertheless interesting to  compare our results with data, and one finds that  the  calculated ``NTcE, pp'' results are in semi-quantitative agreement with the data (with large error bars). In particular for the $\vn{3}$ our model results are slightly higher than data, a discrepancy that would again be expected due to the omitted transverse expansion. Our future study with a realistic hydrodynamic medium will  draw a more definitive answer.

As first pointed out and discussed in details in Refs.~\cite{Zhang:2012mi,Zhang:2012ha}, a hard-soft dihadron correlation would naturally arise due to the hard sector response (via energy loss) and the soft sector response (via collective expansion) to the common initial geometric anisotropy:
\begin{eqnarray}
&& \frac{\int \frac{d\phi_1}{2\pi} \frac{d\phi_2}{2\pi}  2\pi\delta(\phi_2-\phi_1-\Delta\phi) <\frac{dN^{hard}}{dy d\phi_1}\, \frac{dN^{soft}}{dy d\phi_2}>}{<\int \frac{d\phi_1}{2\pi} \frac{dN^{hard}}{dy d\phi_1}>\, <\int \frac{d\phi_2}{2\pi}\frac{dN^{soft}}{dy d\phi_2}>} \ ,  \notag \\
 && \quad \sim \quad 1 + 2 \sum_{n=1,2,3,...} <\vn{n}^s \vn{n}^h > \cos(n\Delta\phi)   \label{eqn:HScorr}
\end{eqnarray}
Properly combining the  high-$\pt$ $\vn{n}$ from energy loss computations and the low-$\pt$ $\vn{n}$ computed in hydro calculations (e.g., \cite{ZhiQiu2011PRC84}), a significant near-side peak is born out in azimuthal angle,  while on the away side a double-hump structure could emerge or not depending on the details of the hierarchy in the $\vn{n}$ vs $n$ spectrum especially with $n=1,\, 2$, and $3$~\cite{Zhang:2012ha,Zhang:2012mi}.   

\section{Possible  jet quenching in high-multiplicity $pPb$ collisions at LHC} \label{sec:ppb}

As briefly discussed in Sec.~\ref{sec:intro}, the nature of high multiplicity events in the LHC pPb collisions (e.g., LHC at $\sqrt{s_{NN}}=5.02$ TeV) is currently under intensive  discussions~\cite{bozek2012prc,Shuryak:2013sra,Dusling:2012wy,Albacete:2013ei,Dumitru:2013tja}: in particular the focal point is whether significant final state interactions occur and whether a collective bulk medium (close to that created in AA collisions) is ever formed in such events.  In Refs.~\cite{bozek2012prc,bozek2013,Bozek:2013df}, the authors have applied hydrodynamic calculations to study the collective expansion of the created matter, assuming its validity for this small system. Their results show sizable elliptical flow   which so far is consistent with the available data \cite{Aad:2013fja,bozek2013}. While it is conceivable that due to the high parton density a certain degree of collective explosion would develop, how such a small system with a very short lifetime could thermalize to the extent of justifying hydrodynamics is quite puzzling, especially provided the  thermalization process is extremely complicated even for the larger and longer AA fireball~\cite{Blaizot:2011xf}. There are also various alternative explanations of measured data based on mostly initial state effects~\cite{Dusling:2012wy,Albacete:2013ei,Dumitru:2013tja}. 

In the present situation, it is of particular interest to find observables other than the soft sector collective expansion that would clearly distinguish the initial state versus final state effects. We here suggest that the hard probe may provide such an opportunity. An extremely dense and (nearly) thermal partonic medium, if indeed formed in high-multiplicity pPb events, will inevitably induce certain amount of final state attenuation to a high energy jet parton traversing the medium. Therefore any measurable effect uniquely from jet energy loss would be a very useful indicator of jet-medium final state interactions. In this section, with the assumption of such a medium in pPb collisions, we explore how large the effects of jet energy loss could be and what would be the best way to detect them,  by applying the same NTcE model used in the corresponding AA study.  

In this calculation, MC calculation is needed, because of the large initial state fluctuations~\cite{bozek2012prc}. In Ref.~\cite{Bzdak:2013zma}, the authors have shown that different implementations of multiplicity generation in the initial nucleon-nucleon scattering, including Glauber and IP-Glassma models \cite{Schenke:2012wb,Schenke:2012hg}, give different initial state eccentricities and different final state elliptical flow. We choose two different energy deposit scenarios in the framework of the MC-Glauber model to generate the initial state \cite{bozek2012prc,bozek2013}: the first assumes a linear relation between entropy density and initial participant density (labeled as ``size a''); the second uses collision density instead (labeled as ``size b''). In general the spacial ``size a'' is bigger than ``size b''. The difference between the two jet quenching results can be used to estimate the uncertainty of our calculations due to the initial state implementation. This is complementary to the study in Ref.~\cite{Bzdak:2013zma}. In Ref.~\cite{bozek2012prc}, the ratio between entropy and participant density, denoted as $s_{0}$ in the following, is estimated to be around $90/\mathrm{fm}$ for both pPb (4.4 TeV) and dPb (3.11 TeV) collisions. In a crude approximation, we use this value in our pPb (5.02 TeV) calculation. 

The NN cross section is $68\ \mathrm{mb}$ at $\sqrt{s_{NN}}=5.02$ TeV \cite{bozek2013}. The distribution of the participant number $N_{p}$ has been checked against the one in Ref.~\cite{bozek2012prc}, based on which we define the most central (i.e. high-multiplicity) events ($0\%-4\%$ centrality) as those with $N_{p}\geq18$. All the following results are shown for this centrality. Moreover since the scattering is not symmetrical with respect to the two colliding beams, the natural ``mid-rapidity'' has been properly shifted ~\cite{ALICE:2012xs}. Although there is no ideal rapidity ``plateau'' in the collision's multiplicity distribution, the data from Ref.~\cite{ALICE:2012xs} indicates that such dependence is  weak. We will assume the boost-invariant longitudinal expansion of the medium: this assumption should be no worse for pPb than for PbPb as in this smaller and shorter-life-time system the transverse expansion is less significant and also the jet's path length would be much shorter. Finally we use the scaling law formula for the pp reference spectrum in this calculation (see related discussion in Sec.~\ref{sec:NTcEmodel}).  

\begin{figure}
\centering
\includegraphics[width=65mm, angle=-90]{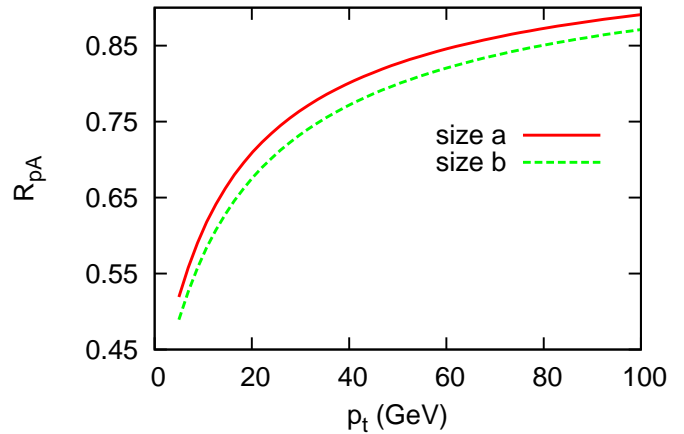}
\caption{$R_{pA}$ vs $\pt$ in the most central pPb collisions.  The ``size a'' and ``size b'' calculations use different initial state implementation. See discussion in the text. } \label{fig:Raappb}
\end{figure}
%
\begin{figure*}
\centering
\includegraphics[height=170mm, width=50mm, angle=-90]{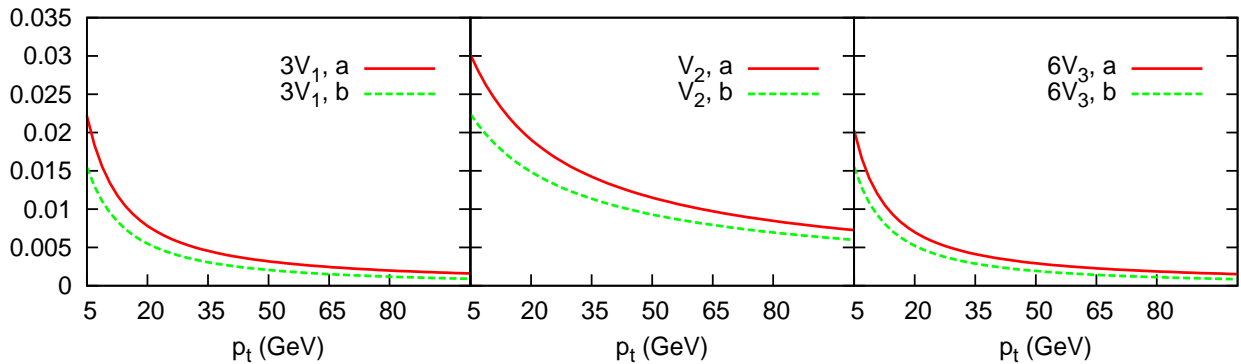}
\caption{$V_{1,2,3}$ vs $\pt$ in the most central pPb collisions. Similar to Fig.~\ref{fig:Raappb}, there are two results (``a'' and ``b'') based on different initial state implementations. $\vn{1}$ and $\vn{3}$ have been rescaled to fit in the plot. Here $\vn{n}$s are the harmonic components projected to the corresponding participant planes. See the discussion in the text.} \label{fig:V123ppb}
\end{figure*}
%

In Fig.~\ref{fig:Raappb}, the computed $\rpa$ (defined in the same way as $\raa$), based on ``size a'' and ``size b'' scenarios, are plotted against the transverse momentum. 
We see that the jet quenching effect is not negligible (particularly in the region below $\sim 20\, \rm GeV$), and is quite close to the $\raa(\pt)$ in the peripheral collisions at LHC (see Fig.~\ref{fig:RAA} for the $30-50\%$ centrality $\raa$ at LHC).
Two important points need to be emphasized here if one were to compare this $\rpa$  directly with the measurements: 1) This suppression is the effect solely from the assumed final state attenuation---there are however possible initial state effects [e.g. from cold nuclear matter (CNM) on Pb side] that may compensate the suppression of $\rpa$; 2) Moreover the pp reference spectrum used here is for the minimum-bias events while it is  likely that for the high multiplicity events the actual initial spectrum from hard collisions could differ considerably from the one used there---most likely being softer thus reducing the suppression effect. Note that the  recently measured $R_{pA}$ by ALICE \cite{ALICE:2012mj} does not show obvious suppression in minimum-bias events, but we focus on the high multiplicity ones, the nature of which however is still unsettled.

Let us now turn to jet quenching azimuthal anisotropy. Different from  the $\rpa$'s interpretation being complicated, the azimuthal anisotropy should be considered as the ``clean'' indicator of the  final state interaction effect.  Fig.~\ref{fig:V123ppb} shows first three harmonics based on ``size a'' and ``b'' initial states. The results already account for the angular difference between $\psih{n}$ and $\psip{n}$. Although the $\vn{2}$ calculated with ``a'' and ``b'' initial states can differ by as large as $\sim 30\%$ (at $\pt=20$ GeV), our results do indicate a significant $\vn{2}\sim 0.01$ in high $\pt$ region---this appears encouraging  enough to be measurable with current luminosity and accuracy at LHC. [One though should be cautious about applying  the current framework to address jets at very high (initial) energy (e.g., 100 GeV) in pPb collisions, because such jets take a major percentage of the system's total energy \cite{Aad:2013fja}.] Meanwhile, $\vn{1}$ and $\vn{3}$ are much less prominent. Recently, ATLAS has measured $\vn{2}$ with $\pt$ up to $4\, \rm GeV$~\cite{Aad:2013fja}. In the collisions with highest total transverse energy (i.e. highest multiplicity), $\vn{2}\sim 0.1$ at $\pt=4$ GeV (based on the two-particle correlation). It is very important and certainly feasible to extend such measurements to higher $\pt$ e.g. $\sim 10\, \rm GeV$.  A measured anisotropy at this magnitude in this high-$p_t$ region would be an unambiguous signal of final state jet-medium interaction and thus also a signal of a dense partonic medium in this ``mini-bang''. 

In addition, similar to the AA situation, combining nonzero high- and low-$\pt$ $\vn{n}$ may produce a hard-soft correlation in pA collisions [see formula~(\ref{eqn:HScorr})]. This correlation can be studied by embedding the jet quenching modeling into the hydrodynamical calculation under current assumption about the medium. On the experimental side,  the corresponding soft-soft correlation in pA has been measured and used to extract the harmonic flows for the soft sector, and it would be of great interest to see the measurements of hard-trigger soft-associate azimuthal correlations in the future.  

\section{Possible jet quenching in high-multiplicity $dAu$ collisions at RHIC} \label{sec:dAu}

\begin{figure}
\centering
\includegraphics[width=65mm, angle=-90]{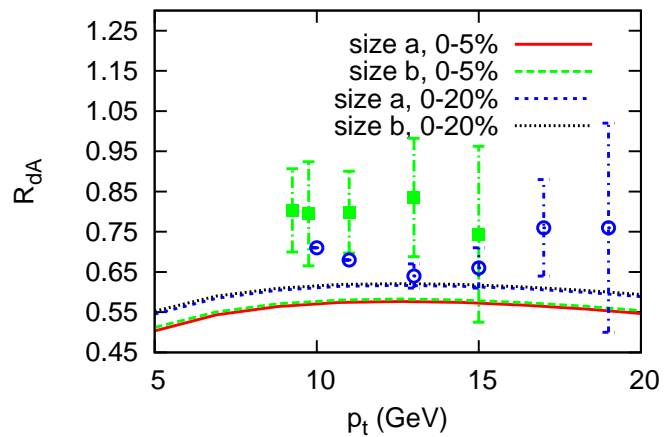}
\caption{$R_{dA}$ vs $\pt$  in dAu collisions. Two different centrality classes, $0-5\%$ and $0-20\%$, are shown. ``size a'' and ``size b'' calculations use different initial state implementations. The data are from PHENIX: (green) squares are $\pi^{0}$'s  $R_{dA}$ measurment \cite{Adler:2006wg} and (blue) circles are $\pi^{0}$'s $R_{cp}$ ($0-20\%/60-88\%$) measurements \cite{Sahlmueller:2012ru,Perepelitsa:2013jua}. Besides the uncertainties shown in the plot, the former has a pp reference normalization uncertainty ($9.7\%$), and the latter has a  collisioin number uncertainty ($8\%$). See discussions in the text.} \label{fig:Raadau}
\end{figure}
\begin{figure*}
\centering
\includegraphics[height=170mm, width=50mm, angle=-90]{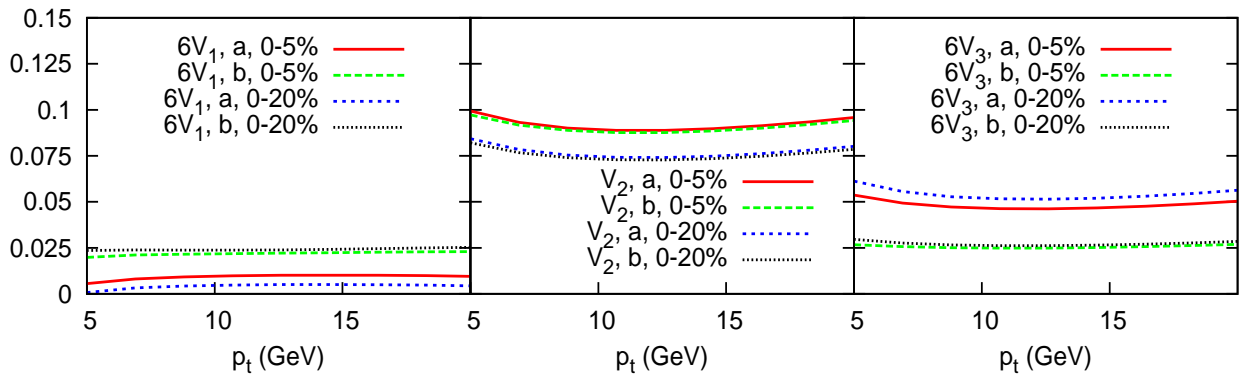}
\caption{$V_{1,2,3}$ vs $\pt$ in dAu collisions. Similar to Fig.~\ref{fig:Raappb}, there are two types of initial state implementations,  ``size a'' and ``size b'', and two centrality classes, $0-5\%$ and $0-20\%$. $\vn{1}$ and $\vn{3}$ have been rescaled to fit in the plot. Here $\vn{n}$s are the harmonic components projected to the corresponding participant planes. See discussions in the text.} \label{fig:V123dau}
\end{figure*}

Motivated by recent observation of the long range correlation in dAu collision (200 GeV) \cite{Adare:2013piz}, we investigate possible jet quenching in these collisions also. The MC-Glauber simulation follows the study of dPb collision (3.11 TeV) in Ref.~\cite{bozek2012prc}. Again we use ``size a'' and ``b''  energy deposit scenarios (see Sec.~\ref{sec:ppb}), and the NTcE model to compute jet energy loss. The total NN cross section, the $p_t$ spectrum in NN collision, and the nucleon density in Au nucleus can be found in the AuAu study in Sec.~\ref{sec:NTcEmodel}. To fix the entropy density to participant density ratio $s_0$ for the dAu collision, we rely on the mid-rapidity multiplicity per participant measurements at RHIC \cite{Back:2003hx} and LHC \cite{ALICE:2012xs}: $dN_{\mathrm{ch}}/\left(d\eta N_p\right)=1.16$ and $2.14$. Based on this, we infer the ratio between $s_0$ in dAu (200 GeV) and in dPb (5.02 TeV) to be $1.16/2.14\sim 0.54$. We then approximate $s_0$ in dPb (5.02 TeV) to be around  $90/\mathrm{fm}$~\cite{bozek2012prc}, and get $s_0=49/\mathrm{fm}$ in the dAu collision.

In Fig.~\ref{fig:Raadau}, we plot the nuclear modification factor $R_{dA}$ ($p_t\geq 5$ GeV) in two different centrality classes, $0-5\%$ with $N_{p} \geq 19$ and $0-20\%$ with $N_{p} \geq 14$. Recent measurements from PHENIX are also shown there, including $\pi^{0}$'s $R_{dA}$ in the $0-20\%$ centrality class \cite{Adler:2006wg} and the ratio ($R_{cp}$) between the $0-20\%$ $R_{dA}$ and the $60-88\%$ one \cite{Sahlmueller:2012ru,Perepelitsa:2013jua}.  Note that the CNM effects mentioned in Sec.~\ref{sec:ppb} also exist in the dAu jet production. Indeed, the mesurement \cite{Sahlmueller:2012ru,Perepelitsa:2013jua} reveals that $60-88\%$ $R_{dA}$  rises to about $1.3$ in the high-$p_t$ region, which is consistent with the CNM effects. 
Here we  expect that the CNM effects' impact on $R_{cp}$ is small and the final state jet attenuation in $60-88\%$ $R_{dA}$ is also negligable. As a result, comparing our result with $R_{cp}$ data is more informative than with $R_{dA}$ ($0-20\%$) data.  
We can see our result agrees reasonably well with the $R_{cp}$ data, although is somewhat lower than the $R_{dA}$ data. 
 
In addition, it is interesting to note that in the $5 \leq p_t\leq 20$ GeV region and $0-5\%$ centrality class, the $R_{dA}$ is smaller than the $R_{pA}$ shown in Fig.~\ref{fig:Raappb}. The differences between the two collisions include: 1) with the impact parameters set as zero, we find that the averaged entropy densities in the center of the cluster(s) are around $100$ and $300$ $\mathrm{fm}^{-3}$ for dAu and pPb respectively; 2) The medium's spacial size in the dAu collision is larger than in the pPb collision; 3) The pp collision's $p_t$ spectrum decreases steeper at $200$ GeV than at $5.02$ TeV; 4) The NTcE effect is reduced from the dAu to the pPb collision. The competion between the first factor and the latter three gives rise to a smaller $R_{dA}$ than $R_{pA}$.

As emphasized in Sec.~\ref{sec:ppb}, high-$p_t$ azimuthal anisotropy observables are important to determine the final state jet attenuation effect. 
In Fig.~\ref{fig:V123dau}, we plot $V_{1,2,3}$ against $p_t$ ($p_t\geq 5$ GeV) in the two different centralities.  We see a large $V_2 \sim 0.1,\, 0.075$ in both two centralities; they are close to the $V_2$ in non-central AuAu collisions (see Fig.~\ref{fig:V2}). This should encourage the corresponding measurements. For $V_{1}$ and $V_3$, ``size a'' and ``b'' results can differ by more than $100\%$. However, the dominance of $V_{2}$ in high-$p_t$ azimuthal anisotropy is clear, which reflects the underlying medium's dipole geometry. 
In addition, the hard-soft correlation for the dAu collision follows the one for the pPb case (see Sec.~\ref{sec:ppb}); thus it is not repeated here.

\section{Summary} \label{sec:sum}

In this paper, we have studied the jet quenching and its azimuthal anisotropy both in AA and high multiplicity pA and dA collisions. We focus on the azimuthal-angle dependent nuclear modification factors in the high-$\pt$ region and investigate how the observable depends on the $p_t$, centrality, collisional beam energies, as well as the different collision systems (AA versus pA and dA). 

For the AA collisions, we have improved  our previous NTcE model studies \cite{Zhang:2012mi,Zhang:2012ie, Zhang:2012ha}, by implementing proper $\pt$ dependence for the mentioned observables [see Eq.~(\ref{eqn:jeteloss1})]. The calculated $\raa(\pt)$ at RHIC ($5\leq\pt\leq 20$ GeV) and LHC ($10\leq\pt\leq 100$ GeV) agrees very well with the data for different centrality classes, in both the $\pt$ dependence and the overall suppression magnitude from $0.2$ to $2.76\, \rm TeV$ collisions. These studies further strengthen the case (as we have illustrated previously) that the NTcE model naturally provides a dynamical reduction of the average opaqueness of the created hot matter from  RHIC to LHC. 

We have further quantified the harmonic coefficients of the $\phi-$dependence of $\raa$ in AA collisions. Our results of the second harmonic, $\vn{2}$, also agree well with the available data at both RHIC ($5\leq\pt\leq 20$ GeV) and LHC ($5\leq\pt\leq 50$ GeV) region. Let us add that, a simultaneous description of both jet quenching and its anisotropy for both RHIC and LHC energies,  has been quite a challenge for many other jet quenching modelings \cite{Jia:2010ee,Zhang:2012ie,Zhang:2012ha,Molnar:2012fn}. Our results for higher order harmonics, $\vn{3}$ and $\vn{4}$, at LHC also show a fair agreement with the CMS measurements  (taking into account the data's large error bars). It does not escape our attention that the   high-$p_t$ $\vn{2}$ and $\vn{3}$ results for LHC AA collisions show the consistent pattern of being   slightly larger than data, which may be attributed to the two factors, i.e., the bulk medium transverse expansion not accounted for in the present modeling, and the mismatch between initial participant planes and the final event planes of the bulk geometry. To fully address these issues it is necessary to combine our energy loss modeling with realistic hydrodynamic medium, which is currently being pursued. 

Finally, motivated by currently intensive discussions on  possibly significant final state interactions  in the high multiplicity pPb events at LHC and dAu events at RHIC, we apply our  NTcE model to explore possible jet quenching effect in these  collisions. We have studied two different initial state implementations in MC-Glauber model. Our $R_{pA}$ and $R_{dA}$ results provide the first quantification of the  ballpark for the magnitude of solely final state suppression effect in such collisions. It is worth pointing out that our $R_{dA}$ at high-$\pt$ is consistent with the recent measurements at RHIC.  We also quantify the high-$\pt$ anisotropy via the harmonic coefficients $\vn{1,2,3,}$, and suggest these to be  a clean signal of possible final state jet-medium interaction. In particular our calculated $\vn{2}$ is on the order of $0.01$ for the pPb collisions, and $0.1$ for the dAu collisions, which should be readily measurable and which, if indeed measured, would be an unambiguous signal of final state jet attenuation in such collisions.

\section*{Acknowledgements}
X.Z. would like to acknowledge the support from US Department of Energy under grant DE-FG02-93ER-40756. J.L. thank the RIKEN BNL Research Center for partial support. We also thank A. M. Sickles for pointing out the dAu measurements and helpful discussions, and B.G. Zakharov for mentioning his recent work on jet quenching in pp collisions.

\end{document}